\documentclass[11pt]{article}
\usepackage{moriond,epsfig}
\usepackage{amsmath}
\usepackage{amssymb}

\def\be{\begin{equation}}
\def\ee{\end{equation}}
\def\bea{\begin{eqnarray}}
\def\eea{\end{eqnarray}}

\begin{document}
\hfill SISSA Ref. 37/2003/EP
\vspace*{3.8cm}
\title{EXTENDED COANNIHILATIONS\\ FROM NON UNIVERSAL SFERMION MASSES}

\author{S.~PROFUMO }

\address{Scuola Internazionale Superiore di Studi Avanzati (SISSA-ISAS)\\ and INFN, Sezione di Trieste,
I-34014 Trieste, Italy}

\maketitle\abstracts{We propose and analyse a GUT-inspired version of the MSSM featuring non universal sfermion masses boundary conditions and asymptotic $b$-$\tau$ Yukawa coupling unification. The particle spectrum of the model gives rise to extended coannihilation processes, involving the lightest sbottom and the tau sneutrino. We study the allowed parameter space, subject to cosmo-phenomenological requirements, and the details of the new coannihilations modes. Implications for direct and indirect neutralino searches, as well as for collider physics, are also briefly summarized.}

\section{Introduction and Motivations}

The physics at the interface of cosmology and elementary particles
theory is currently undergoing an exciting era. On the one side,
we entered an epoch of precision measurements in the field of
astrophysics and cosmology, as demonstrated by the recent results obtained by
the WMAP survey \cite{WMAP}. On the other side, the hunt for
supersymmetry, which would solve a number of open fundamental
questions concerning the physics beyond the standard model, is
going to be boosted both by accelerator \cite{COLLIDERREACH} and by direct or indirect neutralino search
experiments \cite{DIRECT}.

\noindent As regards the theoretical and phenomenological
exploration of supersymmetric models, a number of recent
experimental and observational results increasingly constrain
the candidate scenarios and their parameter space. In this paper
we  deal with the so-called minimal supersymmetric extension
of the standard model, in a \emph{gravity mediated} SUSY breaking
context \cite{SUGRA}. In these models, the numerous \emph{a
priori} free low energy parameters are fixed, through RG running, by the values of the soft SUSY breaking masses and
trilinear terms at some high energy scale
$M_{X}$. The common practice is to assume that some
underlying principle forces these high energy variables to some
\emph{universal} values at $M_{X}$, thus reducing the unkowns to
four numbers and one sign (CMSSM).

\noindent If one further supposes that the electroweak and strong
interactions ultimately unify in a single \emph{grand unification theory}
(GUT), as suggested by the running of the coupling
constants, the GUT gauge group can dictate some relations among
the Yukawa couplings of the high energy theory (Yukawa
Unification) \cite{YU}.

\noindent We will analyse here the phenomenology of a class of
GUT-inspired models, characterized by the relaxation of the
hypothesis of \emph{universality} in the sector of the fermions
scalar superpartners and by partial ($b$-$\tau$) Yukawa Unification
(YU) \cite{SP}. These models are motivated by a minimal $SU(5)$ SUSY GUT
theory in which the scale of grand unification $M_{GUT}$ lies
below the scale $M_{X}$ where universality is
assumed \cite{SUSYGUTNU}. The $SU(5)$ RG running between $M_{X}$ and
$M_{GUT}$ induces a non-universality in the whole scalar sector
(the gaugino sector universality is instead not spoiled by the running \cite{SUSYGUTNU}). As
regards the sfermions, the multiplet structure of the GUT theory,
which collects the particle content into the
$\mathbf{\overline{5}}$ (down-type quark singlet and lepton
doublet) and the $\mathbf{10}$ (up-type quark and lepton singlets
and quark doublet) representations of $SU(5)$, implies the
following parameterization for the soft SUSY breaking masses at
$M_{GUT}$:

\bea & m^2_Q=m^2_U=m^2_E\equiv m^2_{\bf 10}\equiv m^2_0& \\
& m^2_L=m^2_D\equiv m^2_{\bf 5} = K^2m^2_{\bf 10}& \eea

\begin{figure}[!t]
\begin{center}
\includegraphics[scale=0.45,angle=-90]{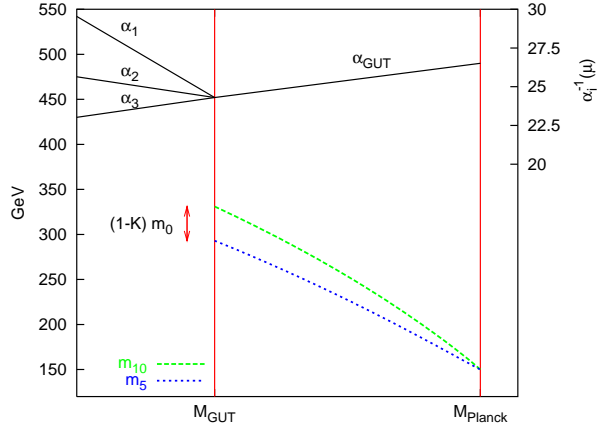}
\end{center}
\caption{The evolution of the coupling constants and of the soft SUSY breaking masses of the ${\bf \overline{5}}$ and ${\bf 10}$ $SU(5)$ multiplets for a $SU(5)$ SUSY GUT theory with $M_{GUT}$ smaller than the scale where scalar universality is assumed (here $M_{\rm Planck}$).}
\label{Nonuniv}
\end{figure}

\noindent In fig.\ref{Nonuniv} we sketch a typical pattern for the
running of $m^2_{\bf 10}$ and $m^2_{\bf 5}$, compared above with
the coupling constants running. Although the full GUT theory would
also imply a large non-universality in the Higgs sector and $K\gtrsim
0.7$, we will resort to a simplification of the model, and let
$m^2_{H_{1,2}}=m^2_0$ in order to concentrate only on the implications of minimal non-universal sfermion masses (mNUSM). We also vary $K$ between 0 and 1 (the latter being the fully universal case). 
 We therefore generically expect a lighter spectrum, with respect to the universal case, for the sparticles belonging
to the $\mathbf{\overline{5}}$ representation for $K<1$.

\noindent On top of  sfermion non-universality, we inherit from the original $SU(5)$ SUSY GUT the
successful prediction of $b$-$\tau$ YU. We impose
exact YU at $M_{GUT}$, and fix the common Yukawa coupling
$h_b=h_\tau(M_{GUT})$ in order to get the right value for
$m_{\tau}(M_Z)$, after having included running effects and the
SUSY thresholds corrections at $M_{SUSY}\equiv\sqrt{m_{\tilde
t_1}m_{\tilde t_2}}$. The output is $m_b(M_Z)$, which is then
required to lie within the properly evolved experimental range
 \cite{MBRANGE}.

\noindent The current upper limit on the cosmological relic
density of neutralinos as cold dark matter candidates requires, in
the CMSSM, efficient suppression mechanisms \cite{STAU,SP}. Among these, a recent
uprise of interest has involved the so-called
\emph{coannihilation} processes \cite{STAU}, occurring when the
next-to-lightest SUSY particle (NLSP) mass is close to the
lightest, which, for cosmological reasons, we require to be the neutralino. The main scope of
the present analysis is to shed some light on the possibility that
the minimal proposed deviation from universality in the sfermion
soft SUSY breaking masses could activate 
\emph{extended coannihilation} channels, involving sparticles belonging to
the $\mathbf{\overline{5}}$ (stau, tau sneutrino, sbottom).
Further, the requirement of $b$-$\tau$ YU, and the fulfillment
of a set of cosmo-phenomenological constraints will strongly
limit the allowed $\tan\beta$ values as well as the sparticles
mass range. The implications of these bounds for accelerator
searches and for direct and indirect SUSY CDM detection will also be
briefly discussed.
\section{The Particle Spectrum}
\begin{figure}[!t]
\begin{center}
\includegraphics[scale=0.45,angle=-90]{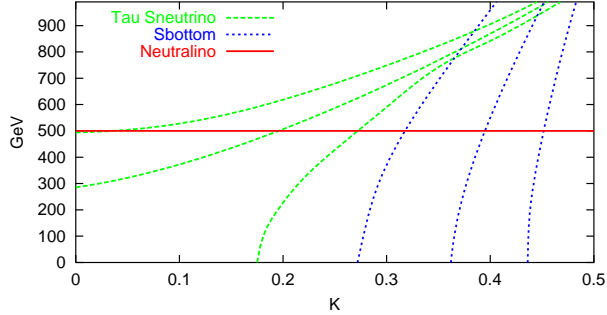}
\end{center}
\caption{The sbottom and sneutrino mass spectrum at $M_{1/2}=1.1\ {\rm TeV}$, $\tan\beta=38.0$ and $A_0=0$, for several values of $m_0$, as functions of the non-universality parameter $K$. For the tau sneutrino, from left to right, $m_0=1350,\ 1650 ,\ 1950$ GeV, while for the sbottom $m_0=2850,\ 3300,\ 3750$ GeV.}
\label{CombSpect}
\end{figure}
\begin{figure}[!b]
\begin{center}
\includegraphics[scale=0.45,angle=-90]{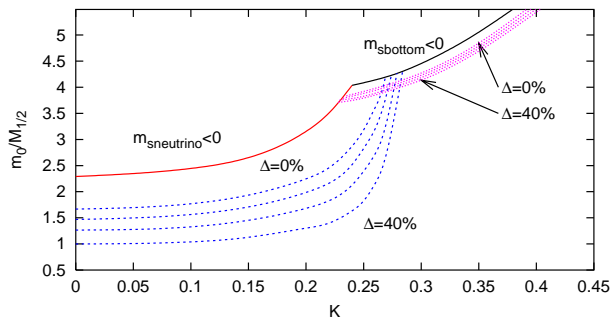}
\end{center}
\caption{Mass isolevel curves for the tau sneutrino and for the sbottom in the $(K,m_0/M_{1/2})$ plane. The solid red and black lines indicate respectively $m_{\tilde\nu_\tau}=0$ and $m_{\tilde b}=0$. The splitting between the isolevel curves is 10\% of the neutralino mass $m_{\tilde\chi}\simeq500\ {\rm GeV}$; $\tan\beta=38.0$, $A_0=0$, $\mu<0$.}
\label{Isolevel}
\end{figure}
The large values of the Yukawa couplings of the third generation
fermions, owing to RG running, yield lighter
superpartners than the ones of the other two
generations. Besides the widely discussed case of the stau
 \cite{STAU}, the role of NLSP, in presence of non-universal
sfermion boundary conditions, can be played by both the sbottom
and the tau sneutrino \cite{SP}. In order to quantify this statement, we
plot in fig.\ref{CombSpect} the masses of the two sparticles at
various values of $m_0$ as functions of the parameter $K$. The features shown in fig.\ref{CombSpect} are
traced back to the generic form of the approximate solutions to
the RG equations at the electroweak scale, which can be
parameterized as
\begin{equation} 
m_{\scriptscriptstyle\tilde\nu,\tilde b}\ \simeq\ m_0\sqrt{a_{\scriptscriptstyle\tilde\nu,  \tilde b}+b_{\scriptscriptstyle\tilde\nu, \tilde b}K^2},
\end{equation}
\noindent where $a$ and $b$ are, to a good approximation, functions of the ratio
$(m_0/M_{1/2})$ and of $\tan\beta$, and weakly depend on the
trilinear coupling $A_0$ \cite{SP}. In all cases we notice that
for a given value of $m_0$ there exists a corresponding range for
$K$ where the sbottom or the tau sneutrino are quasi-degenerate
with the neutralino \footnote{As regards the question of the fine-tuning of the parameter $K$, see the discussion in sec.6.4 of ref. \cite{SP}.} (whose mass roughly depends only on $M_{1/2}$
and is therefore insensitive to $K$). The mass isolevel curves
plotted in fig.\ref{Isolevel} highlight the plausible
coannihilation regions produced by the particle spectrum of NUSM
models (see sec.\ref{sec:coann}). The shape of fig.\ref{Isolevel} is qualitatively
unchanged by the variation of any SUSY parameter, slowly moving to the
right when increasing $\tan\beta$ \cite{SP}. We then conclude that we
expect extended sfermion coannihilations for $K< 0.5$ and for relatively
large values of $m_0\gtrsim1.5\ M_{1/2}$.
\section{Cosmo-Phenomenological Constraints}
\begin{figure}[!t]
\begin{center}
\begin{tabular}{cc}
\includegraphics[width=5.8cm,height=6.6cm,angle=-90]{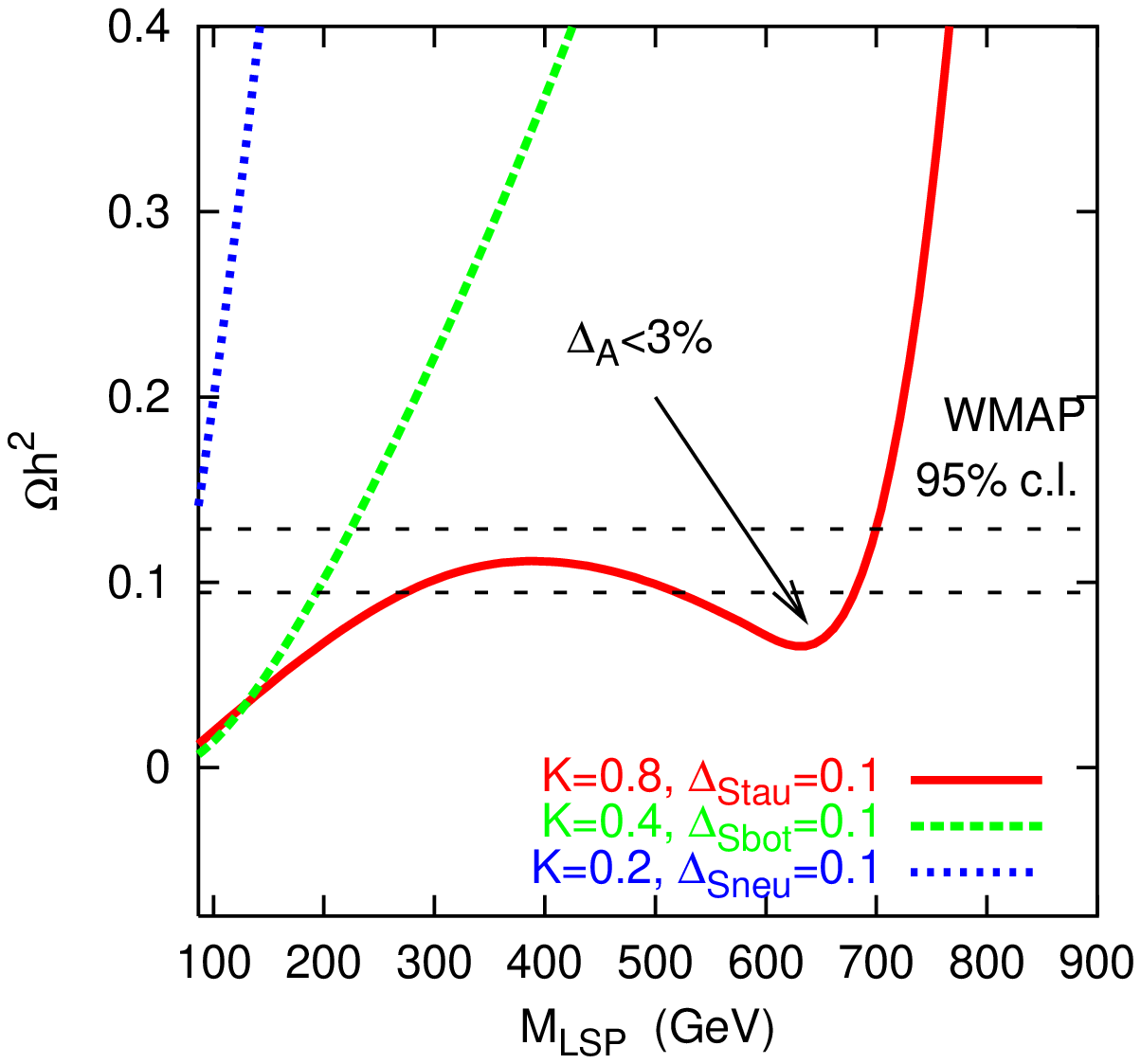} &
\includegraphics[width=6cm,height=6cm,angle=-90]{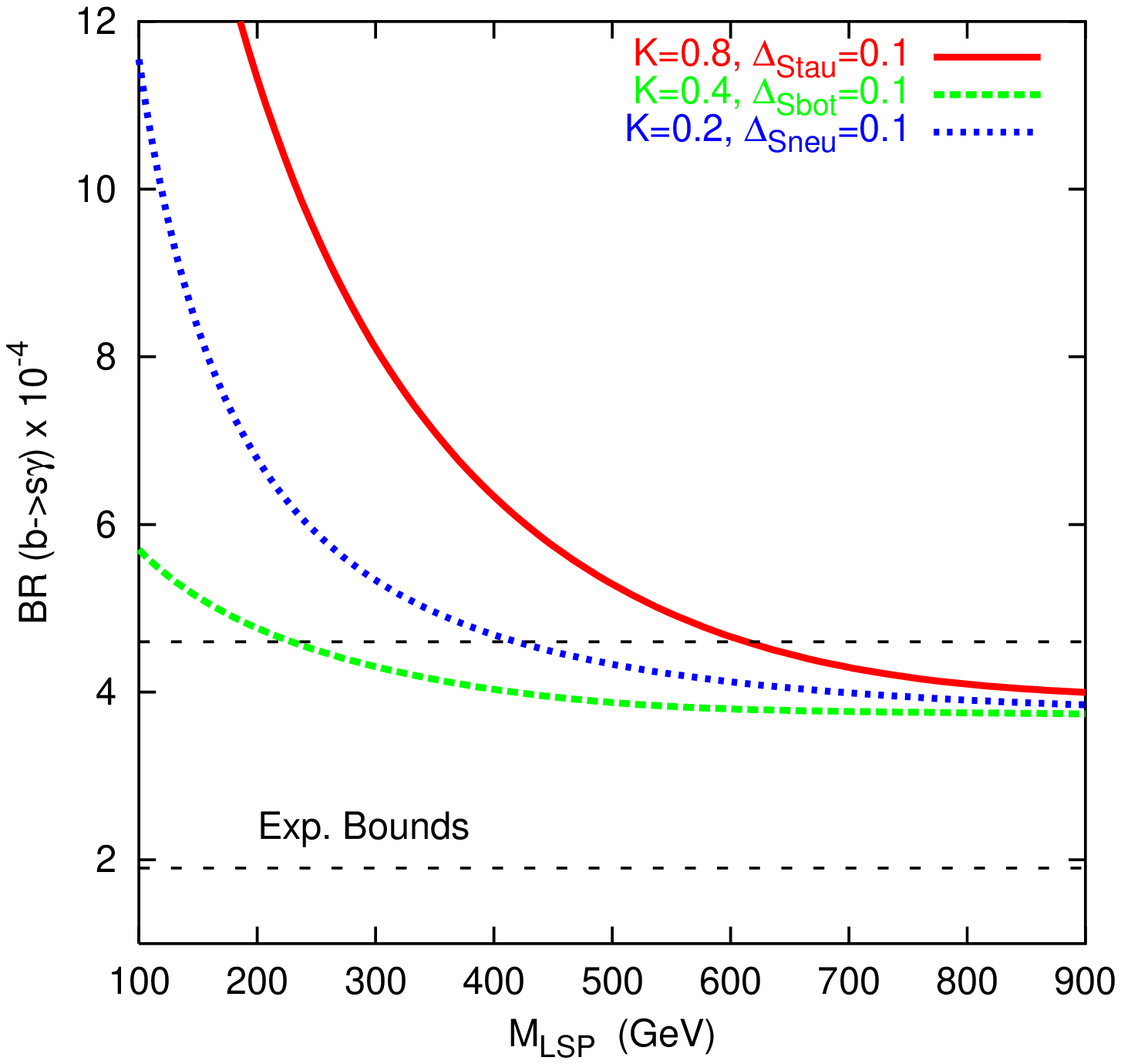}\\
&\\[-0.2cm]
\hspace{1.cm} (\emph{a}) & \hspace{1.cm} (\emph{b})\\
\end{tabular}
\end{center}
\caption{($a$) the neutralino relic density $\Omega_{\tilde\chi}h^2$  and ($b$) the inclusive branching ratio $BR(b\rightarrow s \gamma)$ for three benchmark $K$ cases, corresponding respectively to $NLSP=\tilde\tau,\ \tilde b,\ \tilde\nu_\tau$, at $\tan\beta=36.0$ and fixed $\Delta_{NLSP}=10\%$.}
\label{cosmopheno}
\end{figure}
Supersymmetric models with conserved $R$ parity are commonly
required to fulfill two categories of phenomenological constraints. On
the one side, since these models produce natural candidates for
the CDM content of the Universe, in our case the neutralino, the
corresponding relic abundance as well as the direct or indirect
cosmic detection rates must be compatible with the current
observational and experimental data. On the other side, many
precision accelerator measurements can be used to test the
viability of a given supersymmetric scenario. Finally, the extra
theoretical constraint of $b$-$\tau$ YU forces the mass of the bottom quark to  values to
be compared with the current experimental and lattice data. Details on the numerical calculations and on the statistical
analysis we adopted in this work can be found in ref. \cite{SP}.

\noindent {\bf Neutralino relic density and detection.} We impose
to the model the current determination of the upper limit on
the CDM content of the Universe, coming from the analysis of the
WMAP results \cite{WMAP}, namely $\Omega_{\rm CDM}
h^2=0.1126^{+0.0081}_{-0.0091}$. The lower limit can be evaded
under the hypothesis that other components contribute to the CDM
besides the neutralino. The numerical computation of the neutralino relic density is performed with the updated {\tt micrOMEGAs} code \cite{MICROMEGAS}, which includes thermally averaged exact tree-level cross-sections of all possible (co-)annihilation processes, an appropriate treatment of poles and the one-loop QCD corrections to the Higgs coupling with the fermions. Fig.\ref{cosmopheno} ($a$) shows our results for three {\em benchmark} cases, respectively at $K=0.2,\ 0.4$ and 0.8. In the latter case the coannihilation region overlaps the $A$-pole rapid direct annihilation funnel, extending the allowed $m_{\tilde\chi}$ range up to $\approx700\ {\rm GeV}$. The other two cases, instead, show the typical behavior $\Omega_{\tilde\chi}\propto m^2_{\tilde\chi}$, though suppressed by coannihilation phenomena.\\ 
\noindent As for the direct and indirect neutralino
searches, we verified that in any case the detection rate
estimates for mNUSM models lie below the {\em current} experimental
sensitivity \cite{SP}. We find nonetheless that $t$-channel exchanges of lighter (for $K\ll1$) sfermions generate higher detection rates for small values of $K$, falling within reach of future WIMP search experiments \cite{DIRECT}.

\noindent {\bf Accelerator constraints.} The most stringent bounds
coming from accelerator data are found to be, in the present work,
the inclusive $BR(b\rightarrow s \gamma)$ \cite{BSG} (see fig.\ref{cosmopheno}, frame ($b$)) and the lower
limit on the lightest CP-even Higgs mass \cite{LEP}. They both
give {\em lower} bounds on the mass of the neutralino $m_{\chi}$. As regards the SUSY contributions
to the muon anomalous magnetic moment $\delta a_\mu$,
we chose to resort to the so-called {\em super-conservative}
approach \cite{MW}, and to take as a constraint the extended
5-$\sigma$ bound derived from the latest experimental and theoretical results \cite{DAM} on
$a_\mu$.

\noindent {\bf The $b$ quark mass.} In the context of $b$-$\tau$
YU, the SUSY corrections to the mass of the $b$ quark constrain
the allowed $\tan\beta$ range as well as the sign of $\mu$. In
fact, the sign of these corrections is given by the sign of $\mu$,
and is proportional to $\tan\beta$; since the tree level value of
$m_b$, fixed by the common $b$-$\tau$ Yukawa coupling determined
by $m_\tau$, is typically higher than the upper experimental bound
on $m_b$, the corrections are required to be {\em negative} (hence
$\mu<0$) and {\em large} (hence large $\tan\beta$). In mNUSM models \cite{SP}, the adopted top-down YU excludes $\mu>0$ and
forces, in the $\mu<0$ case, $31<\tan\beta<45$.

\section{Extended Sfermion Coannihilations}\label{sec:coann}
\begin{figure}[!t]
\begin{center}
\begin{tabular}{cc}
\includegraphics[scale=0.35,angle=-90]{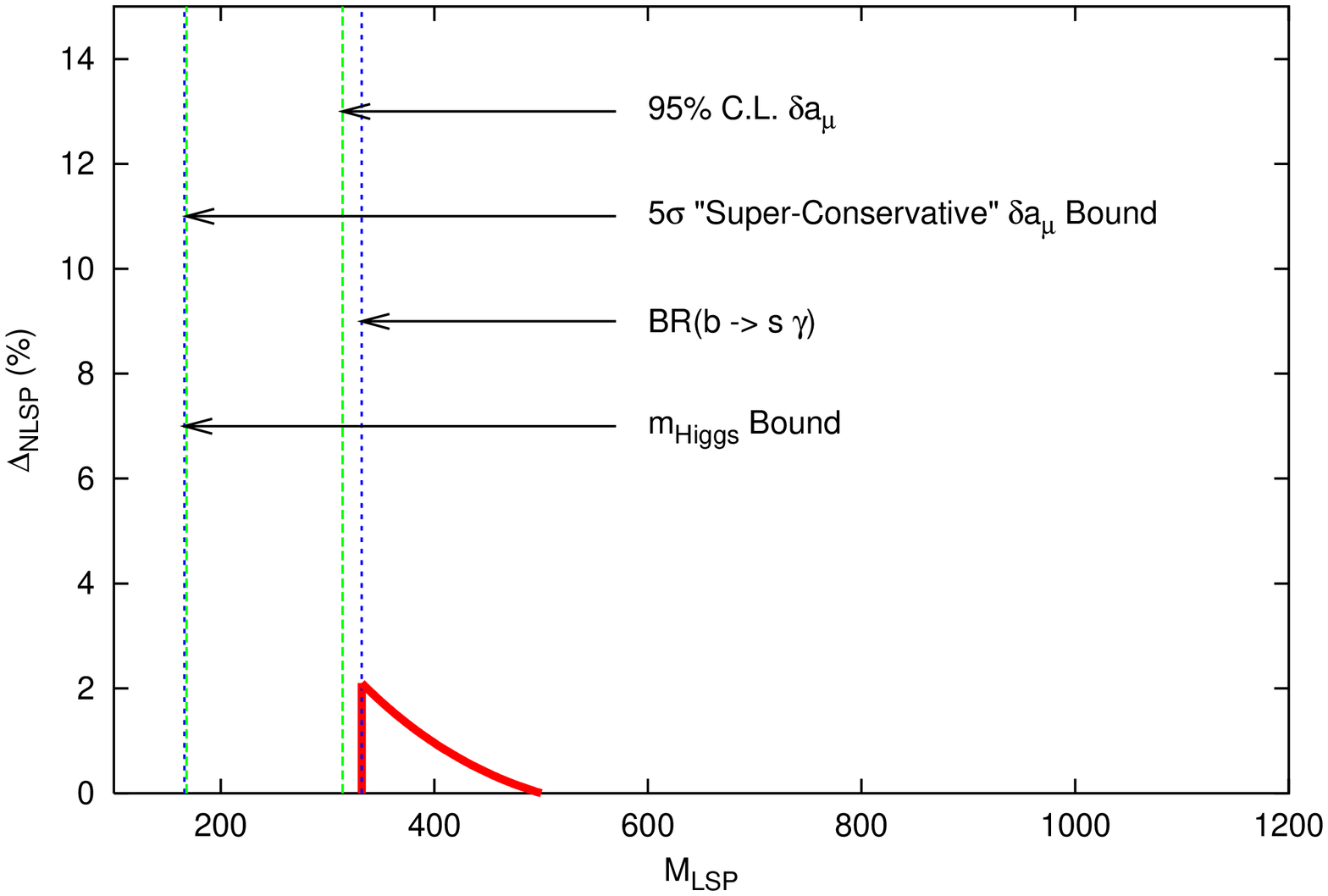} &
\includegraphics[scale=0.35,angle=-90]{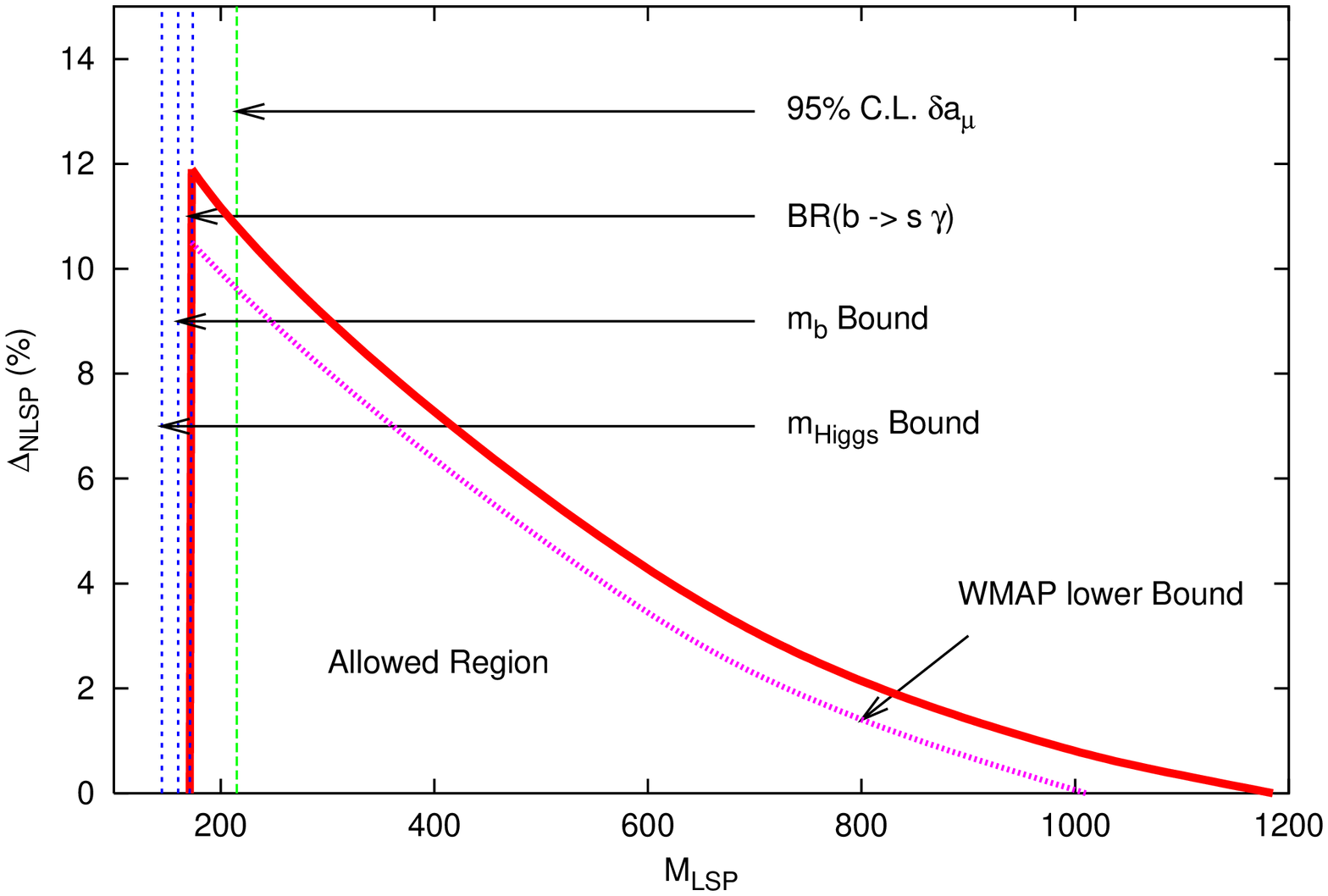}\\
&\\[-0.2cm]
\hspace{1.cm} (\emph{a}) & \hspace{1.cm} (\emph{b})\\
\end{tabular}
\end{center}
\caption{The cosmologically and phenomenologically allowed regions in the ($m_{\tilde\chi},\Delta_{NLSP}$) plane at $\tan\beta=38.0$, at $K=0.1$ ($a$) and $K=0.35$ ($b$).}
\label{coann}
\end{figure}
Minimal NUSM produce a new coannihilation branch at $K<0.5$ (see fig.\ref{Isolevel}), where the coannihilating NLSP is the tau sneutrino at low $1.5\lesssim (m_0/M_{1/2})\lesssim 3$, and the lightest sbottom, for $(m_0/M_{1/2})\gtrsim3$. We study in fig.\ref{coann} the new extended sfermion coannihilations at $A_0=0$ and at an intermediate value of $\tan\beta=38$. Frame ($a$) refers to the tau sneutrino NLSP, at $K=0.1$, while frame ($b$) to the sbottom, at $K=0.35$. We parameterize the remaining variables of the model, $m_0$ and $M_{1/2}$, through $m_{\tilde\chi}$ and $\displaystyle \Delta_{NLSP}\equiv \frac{m_{NLSP}-m_{\tilde\chi}}{m_{\tilde\chi}}$: via RG running, in fact, ($m_0,M_{1/2}$) uniquely determines ($m_{\tilde\chi},\Delta_{NLSP}$), and vice-versa. The NLSP mass range where $\Omega_{\tilde\chi}\lesssim0.13$, at a fixed given $m_{\tilde\chi}$, is as large as allowed by the efficiency of coannihilation processes. We clearly notice that sbottom-neutralino coannihilations (frame ($b$)), which involve strong interaction processes, suppress the relic density much more effectively than in the $\tilde\chi$-$\tilde\nu_\tau$-$\tilde\tau$ case (frame ($a$)). We also find that $\tilde\chi$-$\tilde b$ and $\tilde b$-$\tilde b$ (co-)annihilations evolve, through strong interactions, into few final SM states (respectively $g$-$b$ and $g$-$g$ or $b$-$b^*$) and largely dominate over $\tilde\chi$-$\tilde\chi$ annihilations, while triple $\tilde\chi$-$\tilde\nu_\tau$-$\tilde\tau$ coannihilations present a wider and more complex pattern of final states (see ref. \cite{SP}). In the left part of the figures we show the bounds dictated by the 95\% C.L. exclusion regions of various accelerator constraints. Scanning the values of $\tan\beta$ we find that, for $\tan\beta\gtrsim45$ and $\tan\beta\lesssim31$, the cosmologically allowed region squeezes until no points simultaneously fulfill $b$-$\tau$ YU and the accelerator constraints.

\section{Summary and Conclusions}

We showed that in a $SU(5)$ SUSY GUT with $m_{\bf 5}=Km_{\bf 10}$ scalar soft SUSY masses new extended sfermion coannihilation modes take place. These involve unusual NLSP coannihilating particles, such as the tau sneutrino and the sbottom. The models, to which we apply the constraint of $b$-$\tau$ YU, as inherited from the inspiring $SU(5)$ theory, fulfill the present phenomenological accelerator bounds as well as the recent  cosmological relic density required for the neutralino as cold dark matter constituent.\\
\noindent As regards the detectability of the models under scrutiny, we find \cite{SP} that future direct WIMP search experiments \cite{DIRECT} will be sensitive to part of the highlighted coannihilation regions, namely at low $m_{\tilde\chi}$. Moreover, as far as the CERN LHC $pp$ collider reach is concerned \cite{COLLIDERREACH}, we find that most of the tau-sneutrino coannihilation region will be within reach, while the sbottom coannihilation region (where $m_0\gtrsim 3 M_{1/2}$ and a heavier SUSY spectrum is cosmologically allowed), could escape detectability if $m_{\tilde\chi}\gtrsim 500\ {\rm GeV}$.

\section*{Acknowledgments}
I acknowledge SISSA for financial support and the organizers of the Conference for the very pleasant and stimulating environment.

\end{document}